\documentclass[conference]{IEEEtran}
\IEEEoverridecommandlockouts
\usepackage{cite}
\usepackage{amsmath,amssymb,amsfonts}
\usepackage{algorithmic}
\usepackage{graphicx}
\usepackage{booktabs}
\usepackage{textcomp}
\usepackage{xcolor}
\usepackage{balance}
\usepackage{url}
\usepackage{subcaption}
\usepackage{pdfcomment}
\usepackage{comment}
\usepackage[inline]{enumitem}

\def\BibTeX{{\rm B\kern-.05em{\sc i\kern-.025em b}\kern-.08em
    T\kern-.1667em\lower.7ex\hbox{E}\kern-.125emX}}

\begin{document}

\title{DReyeVR: Democratizing Virtual Reality Driving Simulation for Behavioural $\&$ Interaction Research
}

\author{
\IEEEauthorblockN{Gustavo Silvera \textsuperscript{$\star$} \thanks{\textsuperscript{$\star$} The first two authors contributed equally to this work}}
\IEEEauthorblockA{\textit{Robotics Institute} \\
\textit{Carnegie Mellon University}\\
Pittsburgh, PA, USA \\
gsilvera@andrew.cmu.edu}%
\and
\IEEEauthorblockN{Abhijat Biswas \textsuperscript{$\star$}}
\IEEEauthorblockA{\textit{Robotics Institute} \\
\textit{Carnegie Mellon University}\\
Pittsburgh, PA, USA \\
abhijat@cmu.edu}%
\and
\IEEEauthorblockN{Henny Admoni}
\IEEEauthorblockA{\textit{Robotics Institute} \\
\textit{Carnegie Mellon University}\\
Pittsburgh, PA, USA \\
henny@cmu.edu}%
}
\maketitle

\begin{figure*}[ht]
    \centering
    \begin{subfigure}[t]{0.343\textwidth}
      \includegraphics[width=\textwidth]{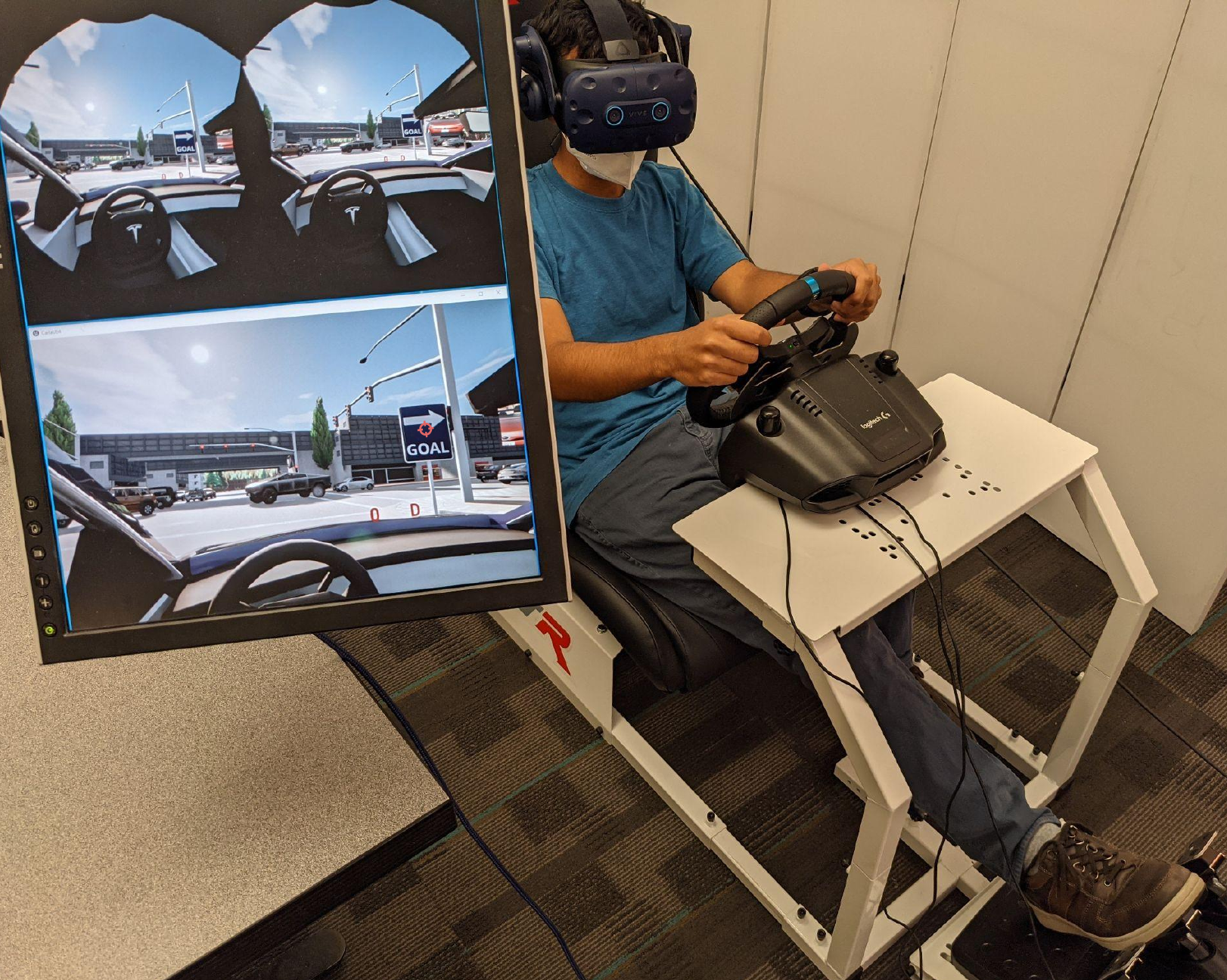}
      \caption{Physical setup with participant driver in driving pose, alongside experimenter's setup monitoring the simulation. Monitor shows the binocular VR view and a flat spectator view showing the gaze reticle. The same scene is shown in more detail on the right.}
      \label{fig:sim-external}
    \end{subfigure}\hfill
    \begin{subfigure}[t]{0.64\textwidth}
      \includegraphics[width=\textwidth]{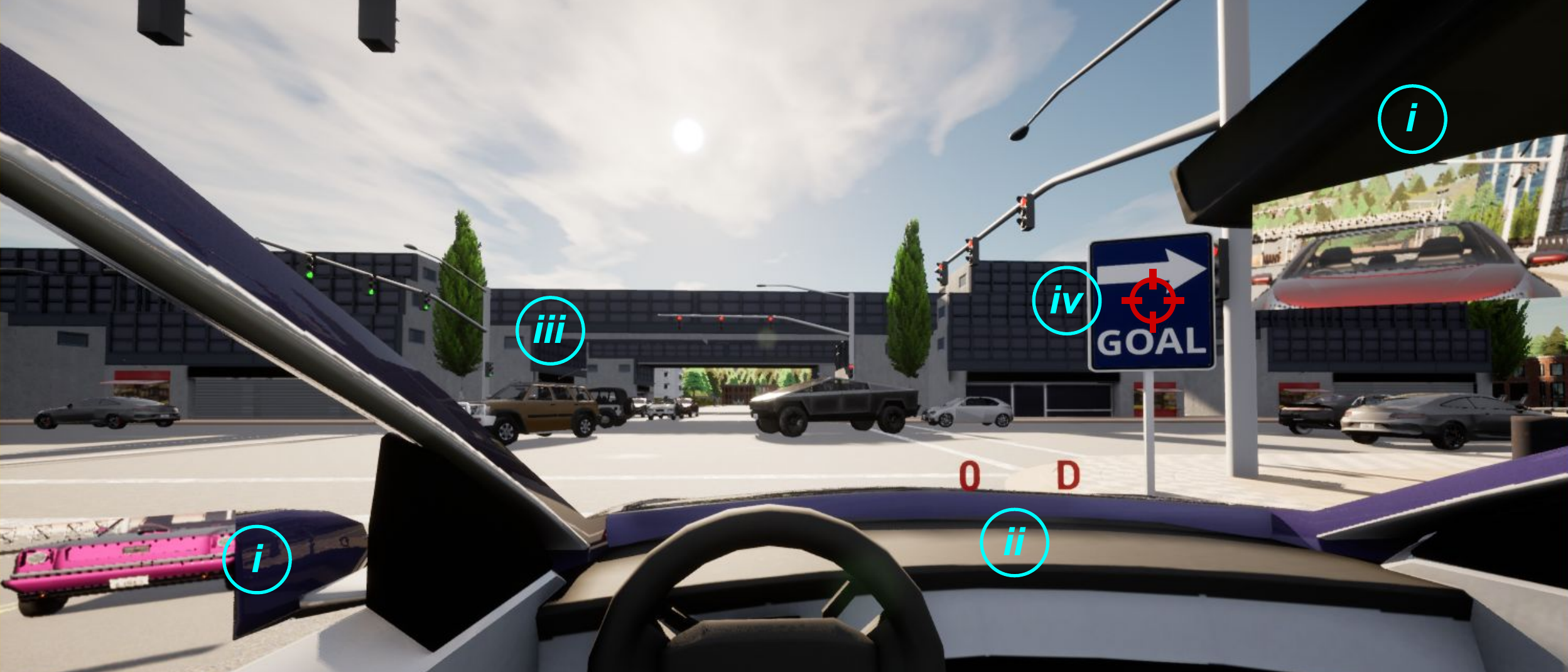}
      \caption{DReyeVR first person perspective during a trial with eye reticle (red crosshair) denoting eye gaze.  
    Some features are highlighted, \textit{i}: rear and side view mirrors added to the ego-vehicle.
    \textit{ii}:  in-vehicle HUD shows speed, engaged gear (Drive/Reverse), turn signal activation (not shown).
    \textit{iii}: custom scenario design is supported, including ego vehicle triggered traffic events, weather controls.
    \textit{iv}: in-world navigational sign props and dynamic placement system is provided.
}
      \label{fig:driver_perspective}
    \end{subfigure}\hfill

    \caption{
    DReyeVR is an open-source VR-based driving simulation platform for behavioural research. We both describe a viable hardware design using consumer products and provide the requisite open-source software to run human subjects studies.
    }
    \label{fig:overview}
\end{figure*}

\begin{abstract}

Simulators are an essential tool for behavioural and interaction research on driving, due to the safety, cost, and experimental control issues of on-road driving experiments. The most advanced simulators use expensive 360 degree projections systems to ensure visual fidelity, full field of view, and immersion. However, similar visual fidelity can be achieved affordably using a virtual reality (VR) based visual interface. We present DReyeVR, an open-source VR based driving simulator platform designed with behavioural and interaction research priorities in mind. DReyeVR (read ``driver'') is based on Unreal Engine and the CARLA autonomous vehicle simulator and has features such as eye tracking, a functional driving heads-up display (HUD) and vehicle audio, custom definable routes and traffic scenarios, experimental logging, replay capabilities, and compatibility with ROS. We describe the hardware required to deploy this simulator for under $5000$ USD, much cheaper than commercially available simulators. Finally, we describe how DReyeVR may be leveraged to answer an interaction research question in an example scenario. 
DReyeVR is open-source at this \href{https://github.com/HARPLab/DReyeVR}{url}. \footnote{This work was funded in part by the National Science Foundation (IIS-1900821).} 

\end{abstract}


\section{Introduction}



Emerging intelligent systems in vehicles, from Advanced Driver Assistance Systems (ADAS) to the upper levels of SAE-defined vehicular autonomy~\cite{sae2014taxonomy}, is turning vehicles into autonomous agents in their own right. To enable safe and efficient driving experiences alongside such autonomy, researchers are studying various aspects of driver behaviour and driver-vehicle interaction (see \cite{review2021Behavioral} for review). For instance, researchers have sought
to understand how much time it takes for drivers to gain situational awareness after a period of inattentiveness~\cite{lu2017much} or if the safety of handovers from autonomous to manual driving in critical situations can be predicted from driver eye gaze preceding the handover~\cite{wu2019gaze}.

Usually, carrying out such studies in the real world is not just expensive and/or dangerous, it also leads to immense challenges in measurement, instrumentation, and environmental control. 
Hence, driving simulations play a valuable role in collecting human data for behavioural and interaction research. 


Advances in computer graphics and  projection systems have resulted in high visual fidelity simulators with real vehicle cabs surrounded by $360^{\circ}$ Cave Automatic Virtual Environments (CAVEs). However, due to high cost ($\sim300,000$ USD), simulators with these visual interfaces remain inaccessible to most researchers, finding use primarily in industrial or national government labs~\cite{haug1990nads, jamson2007uolds, zeeb2010daimler}. 

Instead, researchers commonly use a series of joined flat screens as the visual interface, which comes at the cost of a reduced driver Field of View (FoV) and immersion~\cite{konstantopoulos2010driver, crundall2012some}. Some groups have attempted to circumvent the cost of CAVE visual systems while maintaining $360^{\circ}$ FoV by using consumer virtual reality (VR) headsets~\cite{taheri2017virtual, goedicke2018vr, xu2020much}.
Apart from the low cost, previous studies have found that for a simulated driving task, user concentration, involvement, immersion and enjoyment was greater when drivers experienced the simulator via virtual reality rather than on a flat screen \cite{comparing2018Lhemedu, Comparing2018Pallavicini}. VR-based simulators also allow for accurate head pose and eye gaze estimation without restricting FoV or naturalistic head or eye movements --- important signals for behavioural research.

Virtual reality is an increasingly popular modality in human-robot interaction research~\cite{williams2019reality}, seeing the emergence of new VR robotics tools \cite{whitney2018ros} and use in automobile driving~\cite{goedicke2018vr, xu2020much}, motorized boat driving~\cite{novitzky2020virtual}, wheelchair driving~\cite{vailland2020vestibular} among other areas including manipulation~\cite{duguleana2011evaluating}, teleoperation~\cite{naceri2019towards, taylor2020diminished}, and deictic communication~\cite{williams2018augmented}. 




In this work, we introduce our open-source VR based driving simulation platform for behavioural and interaction research involving human drivers: DReyeVR. DReyeVR features support for collecting signals that are widely used in behavioural driving research including eye gaze and head pose. It also supports researchers by providing an experimental monitoring and logging system to record and replay trials as well as a sign-based navigation system to direct drivers in the simulator. We build atop CARLA \cite{carla}, a popular open source driving simulator for autonomous driving, and Unreal Engine 4 (UE4) adding naturalistic visual (\textit{e.g.} in-vehicle mirrors) and auditory (\textit{e.g.} vehicular $\&$ ambient sounds) interfaces allowing ecologically valid information gathering behaviour by drivers. Our foundations provide access to semantic information about the simulated world, and a flexible interface to define custom traffic scenarios with granular control over other vehicles, pedestrians, and traffic lights --- all useful experimental tools for researchers. 

In the rest of the paper, we provide a detailed description of DReyeVR's features and software stack. We also describe the hardware required to establish a workable driving simulator using DReyeVR for $<5000$ USD and example usage. 




\section{Related work}
Driving simulation is a widely used tool which allows for tightly controlled study of driver behaviour, especially in dangerous situations which cannot be studied on-road due to safety and ethical concerns. These tools comprise a range of physical, visual, and motion interfaces which have been validated to different degrees (see \cite{slob2008state, caird2011twelve, wynne2019systematic} for review). In most cases, behavioural trends tend to transfer from driving simulation to on-road driving even if only in relative terms rather than absolute numerical effect sizes \cite{caird2011twelve}.

\textbf{Driving simulators: Hardware design} The most realistic simulators ($>12$ on the simulator fidelity scale \cite{wynne2019systematic}) use CAVE $360^\circ$ projection systems, physical vehicle cabs, and Stewart platforms for motion realism \cite{jamson2007uolds, haug1990nads, zeeb2010daimler, artz2001design}. These usually require resources that are only feasible on the scale of national government or industry projects and are hence inaccessible to the majority of academic labs. 

Virtual reality (VR) based simulators offer an alternative that is orders of magnitude cheaper to CAVE visual systems (a few thousand USD vs hundreds of thousands of USD) while still maintaining an immersive $360^\circ$ FoV. Promisingly, \cite{pai2019promise} found that hazard perception related effects discovered in a high fidelity simulator with a vehicle cab were replicable in a VR based simulator.


\textbf{Driving simulators: Software design}
Behavioural studies that use driving simulators (in VR or otherwise) usually construct them ad-hoc for their particular needs and research questions, resulting in closed-source driving simulators and repeated development cost across groups. 

On the other hand, CARLA \cite{carla} is an open-source simulator for autonomous driving development and research, among others (AirSim \cite{shah2018airsim}, DeepDrive\cite{deepdrive}, TORCS \cite{wymann2000torcs}). These all focus on training and evaluating algorithmic and autonomous driving with little emphasis on human driving. Consequently, none of these simulators natively support or prioritize VR development as this is irrelevant for algorithmic agent use. The closest alternative to DReyeVR would be \cite{michalik2021developing}, which is an independent driving simulator for behavioral research. However, it prioritizes a traditional flat-screen display interface over VR, which limits naturalistic eye gaze and head movement behaviour and measurement. They also do not provide the flexibility of user defined traffic scenarios and a recording/replay system that DReyeVR does through its relationship with CARLA and do not open-source their simulator.

\section{DReyeVR Features}

Emphasizing human interaction, DReyeVR builds on top of CARLA \cite{carla} to add novel features necessary for behavioural and interaction research. Our contributed features are:
\begin{enumerate*}
    \item \textbf{Ego vehicle control}: We extend CARLA's flat-screen ego vehicle controller to an immersive and maneuverable VR interface between a human driver and the world. 
    \item \textbf{Sound design}: CARLA is an audio-less simulator designed for visual AI drivers, but to enhance immersion for human drivers, we add spatially accurate sound cues in DReyeVR.
    \item \textbf{In-environment directions}: 
    We provide a set of in-world directional road signs as a human-readable navigation system and a means for dynamically placing them at runtime  (Fig.~\ref{fig:driver_perspective}\textit{iv}).
    \item \textbf{Eye tracking and recording/replaying}: We create a custom virtual sensor for collecting eye gaze, head pose, and human inputs that can be recorded and replayed. 
    \item \textbf{Experimental monitoring system}:  We include an interface for monitoring the progress and head/eye gaze of a participant during an experiment.
    \item \textbf{\texttt{ScenarioRunner} integration}: We provide integration with CARLA's traffic scenario management tool, \texttt{ScenarioRunner}, to operate with DReyeVR's custom ego-vehicle and sensor.
\end{enumerate*}

\subsection{Ego Vehicle}
CARLA's built-in vehicles are not designed for human drivers and therefore do not include several essential features that humans often rely on while driving. For instance, drivers depend on their vehicle's side and rear-view mirrors to quickly gather information about their surroundings and blind spots.



To support this, we added side and rear-view mirrors as an additional stream of information allowing drivers to maintain environmental awareness in a naturalistic manner (Fig.\ref{fig:driver_perspective}\textit{i}).

We also added a heads-up display inside the vehicle's dashboard containing a digital speedometer, drive/reverse indicators, and turn signals (Fig.~\ref{fig:driver_perspective}\textit{ii}) for vehicular monitoring. 

Finally, to maneuver the vehicle, we provide integration with external hardware that mimics the posture and inputs of realistic driving. This hardware includes an arcade seat, steering wheel with force-feedback, turn signal paddles, and brake/accelerator pedals (Fig.~\ref{fig:sim-external}). It is still possible for the ego-vehicle to be algorithmically controlled and also to handoff control to the participant, allowing simulation involving partially and fully autonomous vehicles. 

\subsection{Sound Design}
World audio is absent from the CARLA simulator, but is another important aspect of driving simulation for behavioural research since humans  use and  spatial sound cues to  track information such as the location of other vehicles, the engine gear, state of their own vehicle etc.

In DReyeVR, we give each vehicle its own responsive engine revving sound and pedestrians walking/talking sounds, allowing for an attentive human driver to make use of this information and locate other actors around them. Additionally, the engine revving sound is parameterized by the engine RPM and location, allowing drivers to infer the speeds and accelerations of surrounding vehicles. 
We also added other ego-vehicle only sounds for sensory information on driver inputs, such as turn signal clicks and gear shifts.
Further, we added a spatially accurate and velocity-dependent collision sound generator, allowing drivers to infer the intensity and location of a collision in the absence of physical motion cues.

To complete the environmental sound design, we added relevant sound to various regions in the CARLA maps: \textit{e.g.} areas with plentiful vegetation have bird chirping sounds, and city/underground regions have wind and traffic/urban sounds.

\subsection{In-Environment Directions}
\label{sec:inenv_directions}

CARLA's primary method of providing directional instructions to its users (autonomous agents) is via a \texttt{.json} file describing the entire route through waypoint coordinates --- not a suitable interface for humans. Instead, road signs are a natural environmental cue for navigation that modify driver eye gaze in an ecologically valid manner. For instance, ``Right to goal'' signs can be placed at an intersection (Fig.~\ref{fig:driver_perspective}\textit{iv}). We added a system allowing these to be pre-specified during experiment design (see documentation) and then dynamically placed at runtime. This allows scenario designs (and corresponding sign placements) with different routes on the same map to be swapped easily during experiments.

\subsection{Eye Tracking}
To integrate eye tracking, an important signal in behavioural research, we elected to use an HTC Vive Pro Eye as the head-mounted device for DReyeVR because of its built-in eye tracking and head pose tracking sensors. Using HTC's proprietary developer SDK for their VR eye tracker (SRanipal), we integrated their UE4 plugin with our system to passively track and record eye tracker data in real time. This provides information on individual eye gaze directions, pupil diameters and positions, and eye openness. These are important signals that can be used to monitor the driver's attentiveness, cognitive load, and infer their intent.

To collect this information, we created a custom CARLA sensor that tracks eye gaze data through SRanipal, head pose data through SteamVR (UE4 plugin), and all other user inputs through the corresponding hardware interface. The sensor also can easily stream data to the CARLA client API. Asynchronous eye tracker querying can be employed to collect data untethered from CARLA's simulator tick, our headset supports up to $120$Hz eye tracking. 


\subsection{Experimental Recorder and Replayer}


Simulation research provides experimenters the useful ability to effortlessly record and replay previous simulator data for further analysis. CARLA includes a native recorder and replayer system that serializes world information on every simulator tick for post simulation reenactment, but this is only meant to track world actors (doesn't include ego-vehicle or sensors such as eye tracker). 

Alongside the default CARLA world state recording, recording and replaying the human driver's sensory and mechanical inputs is important for behavioural research as it allows post-hoc analysis and precise comparison between different participants. We added support for recording custom sensor data to CARLA's binary recorder file, and converting this data back to a human-readable format when desired. Similarly, replaying the DReyeVR sensor data supports visualizing user gaze and head pose data accurately, so an experimenter understands exactly where the driver was looking at any time.

Additionally, CARLA only provides simple controls for controlling the replay status through an external python API, which is tedious to use interactively. To make this a more natural interface, we added support to play, pause, restart, speed up/down, and skip ahead in the replay, all within the simulator itself. These ``media controls'' allow experimenters to easily control the replay status on demand, which is useful for studying interesting sections in the recording. 

\textbf{ROS compatibility} We extend CARLA's ROS integration to include our ego-vehicle and sensor data, which is useful for users who want to test online driving assistance algorithms (\textit{e.g.} an eye gaze and world state based hazard warning system) without modifying the DReyeVR C++ simulator code.

\subsection{Custom Traffic Scenario Support}
Within the CARLA ecosystem, \texttt{ScenarioRunner}, a behaviour tree based action model, is used to design and execute ``scenarios'' with dynamic triggers and environmental effects for vehicles to navigate. CARLA contains a library of (NHTSA recommended~\cite{najm2007pre}) pre-crash scenarios.
DReyeVR extends compatibility with existing \texttt{ScenarioRunner} scenarios and features such as scoring our driver based on their route performance, starting and stopping a recording/replay, and all other client vehicle commands.

\section{System implementation and Usage} 

\begin{figure}[htbp]
    \centering
    \includegraphics[width=0.95\textwidth / 2]{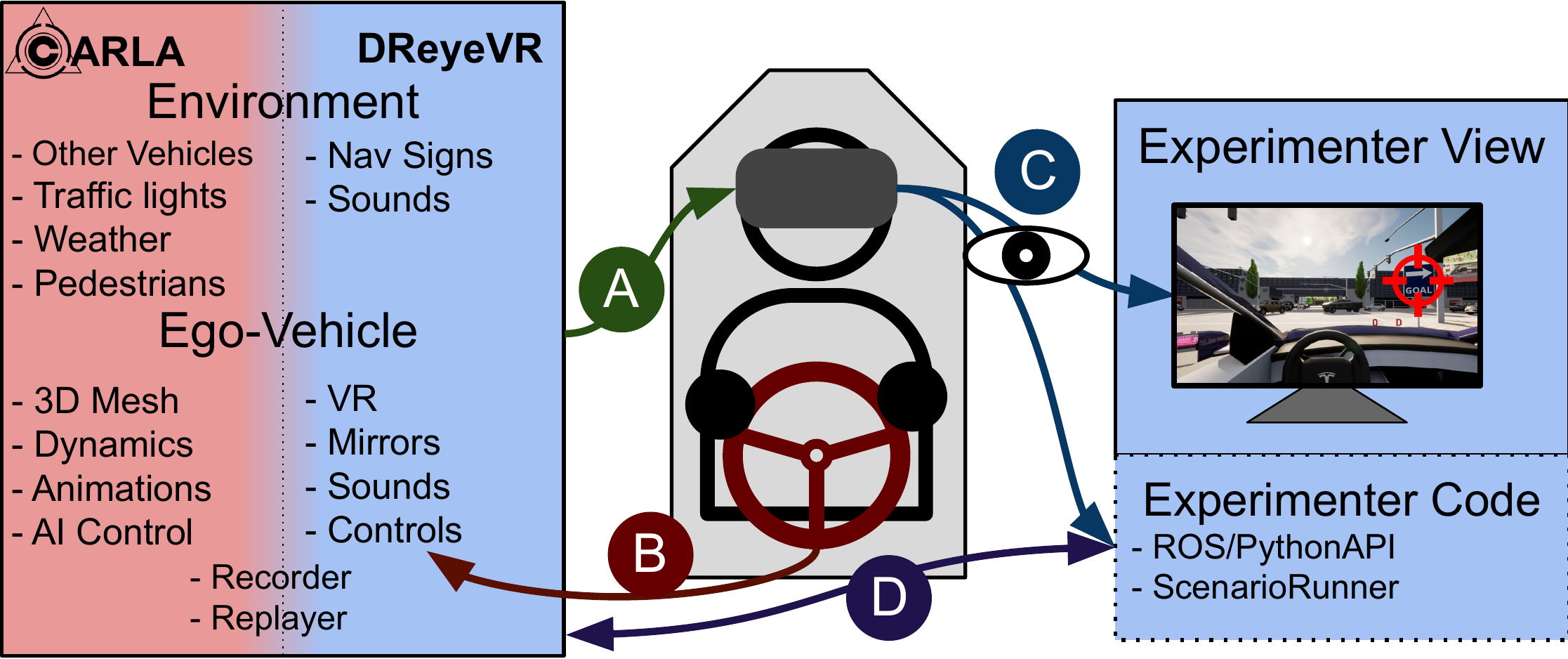}
    \caption{System schematics denoting DReyeVR (blue) and CARLA (red) alongside signals such as A: VR environment and vehicle to driver. B: Human driver inputs. C: Head pose and eye tracker data. D: DReyeVR data streaming to client.}
    \label{fig:systems_testing}
\end{figure}

\textbf{Hardware setup:} During testing, we used the HTC Vive Pro Eye as our head-mounted device (HMD) which has SteamVR support, built-in eye tracking, and an available eye tracking SDK. The kit included $2$ Valve base (tracking) stations, and two controllers. For our driving hardware we elected to use a Logitech G29 wheel and pedals kit, which has an open source UE4 plugin, mounted to a GTR arcade seat for comfort and improved immersion. DReyeVR is modular, and uses macros to enable HMD and racing wheel specific features at runtime so a different HMD or racing wheel can be used with some changes (replacing with the appropriate UE4 plugin). 
This also extends to using other sensors (say a 3rd-party gaze tracker or a pulse oximeter to measure cognitive load) alongside our simulator, which can be interfaced directly with UE4 or can use the ROS compatibility.


\textbf{Software and computation:}
DReyeVR is currently dependent on the Windows 10 OS for proprietary SDKs such as HTC's SRanipal eye tracker driver. While Windows is recommended for optimized VR support, all our work translates to Linux systems except for the eye tracking and hardware integration which have Windows-only dependencies. We have tested that it is possible to connect a CARLA ROS bridge running on a Linux machine to a CARLA server instance running on a Windows machine. This enables streaming data from the host simulator, including all DReyeVR sensor data, to a ROS core on Linux with minimal latency. 



\textbf{DReyeVR Simulator fidelity:}
According to the simulator fidelity scoring system developed by the recent meta-review of Wynne \textit{et al.}, DReyeVR scores $9/15$ points, ranking us in the medium simulator fidelity category \cite{wynne2019systematic}. The score breakdown is: 1 (Motion, No motion base) + 5 (Visual, $>270^\circ$ FOV) + 3 (Physical, Arcade seat with steering wheel) = 9 (\textit{medium} simulator fidelity).
For comparison, a similar setup with conjoined flat screens instead of VR \cite{crundall2012some} would scored 6 points, a fixed-base simulator with a vehicle cab and CAVE visual system would score 11 and one with a full 6-DOF motion base would score 15 \cite{zeeb2010daimler}. 




\textbf{Example usage:}
Previous research has explored how eye gaze can be used to track situational awareness (SA) across task relevant stimuli in a scene [1]. However, once such a system predicts a low awareness of a task relevant stimulus, advanced driver-assistance systems (ADAS) must intervene in some manner to prevent a dangerous situation such as by alerting the driver to enhance SA or via a direct driving intervention. We seek to investigate the effectiveness and perceived safety of these assistive driving interventions in comparison to each other. We do so by designing a between subjects simulated driving study in which the independent variable is the intervention method used by the ADAS in a safety critical driving situation. In our designed situation, a pedestrian emerges from between two parked cars and starts jaywalking. It is possible to spot the pedestrian a few metres ahead of the scenario starting but our pilots indicate that very few people actually spot it if the scenario happens just after a busy intersection.
 There are three conditions: Alert, Highlight, and Takeover. In Alert the ADAS makes a sound to indicate that a task relevant stimulus has been missed by the driver. In Highlight, the ADAS uses the simulated windshield to (similar to augmented reality) highlight the task relevant stimulus. Finally, in Takeover, the ADAS takes over and applies the brakes in order to prevent collision with the jaywalker.

Our simulator, crucially, allows us to ask participants to deal with dangerous driving situations and use objective metrics (such as collisions and reaction times) to measure driving quality without actual danger. Additionally, DReyeVR enables custom scenarios and different types of interventions to all be programmed into the same simulator while also tracking eye gaze in real time and providing a real-time API.

\section{Conclusion}

We present DReyeVR, an open-source VR based driving simulation platform, designed especially for use for behavioural and interaction research involving manual, partially autonomous, and fully autonomous driving. DReyeVR adapts the CARLA autonomous driving simulator adding crucial features for human use while maintaining compatibility with CARLA systems allowing flexible control of the environment around the ego-vehicle. 
DReyeVR is well documented with guides for installation, usage, and extending with custom sensors and is available at this \href{https://github.com/HARPLab/DReyeVR}{url}. 

\bibliographystyle{IEEEtran}
\balance
\bibliography{refs}

\begin{thebibliography}{10}
\providecommand{\url}[1]{#1}
\csname url@samestyle\endcsname
\providecommand{\newblock}{\relax}
\providecommand{\bibinfo}[2]{#2}
\providecommand{\BIBentrySTDinterwordspacing}{\spaceskip=0pt\relax}
\providecommand{\BIBentryALTinterwordstretchfactor}{4}
\providecommand{\BIBentryALTinterwordspacing}{\spaceskip=\fontdimen2\font plus
\BIBentryALTinterwordstretchfactor\fontdimen3\font minus
  \fontdimen4\font\relax}
\providecommand{\BIBforeignlanguage}[2]{{%
\expandafter\ifx\csname l@#1\endcsname\relax
\typeout{** WARNING: IEEEtran.bst: No hyphenation pattern has been}%
\typeout{** loaded for the language `#1'. Using the pattern for}%
\typeout{** the default language instead.}%
\else
\language=\csname l@#1\endcsname
\fi
#2}}
\providecommand{\BIBdecl}{\relax}
\BIBdecl

\bibitem{sae2014taxonomy}
S.~O.-R. A. V.~S. Committee \emph{et~al.}, ``Taxonomy and definitions for terms
  related to on-road motor vehicle automated driving systems,'' \emph{SAE
  Standard J}, vol. 3016, pp. 1--16, 2014.

\bibitem{review2021Behavioral}
I.~Kotseruba and J.~K. Tsotsos, ``Behavioral research and practical
  applications of drivers' attention,'' \emph{arXiv:2104.05677}, 2021.

\bibitem{lu2017much}
Z.~Lu, X.~Coster, and J.~De~Winter, ``How much time do drivers need to obtain
  situation awareness? a laboratory-based study of automated driving,''
  \emph{Applied ergonomics}, vol.~60, pp. 293--304, 2017.

\bibitem{wu2019gaze}
M.~Wu, T.~Louw, M.~Lahijanian, W.~Ruan, X.~Huang, N.~Merat, and M.~Kwiatkowska,
  ``Gaze-based intention anticipation over driving manoeuvres in
  semi-autonomous vehicles,'' in \emph{2019 IEEE/RSJ International Conference
  on Intelligent Robots and Systems (IROS)}.\hskip 1em plus 0.5em minus
  0.4em\relax IEEE, 2019, pp. 6210--6216.

\bibitem{haug1990nads}
E.~J. Haug, ``Feasibility study and conceptual design of a national advanced
  driving simulator. final report,'' Tech. Rep., 1990.

\bibitem{jamson2007uolds}
A.~H. Jamson, A.~J. Horrobin, and R.~A. Auckland, ``Whatever happened to the
  lads? design and development of the new university of leeds driving
  simulator,'' in \emph{Proceedings of the Driving Simulation Conference,
  Driving Simulation Association}.\hskip 1em plus 0.5em minus 0.4em\relax
  Citeseer, 2007.

\bibitem{zeeb2010daimler}
E.~Zeeb, ``Daimler’s new full-scale, high-dynamic driving simulator--a
  technical overview,'' \emph{Actes INRETS}, pp. 157--165, 2010.

\bibitem{konstantopoulos2010driver}
P.~Konstantopoulos, P.~Chapman, and D.~Crundall, ``Driver's visual attention as
  a function of driving experience and visibility. using a driving simulator to
  explore drivers’ eye movements in day, night and rain driving,''
  \emph{Accident Analysis \& Prevention}, vol.~42, no.~3, pp. 827--834, 2010.

\bibitem{crundall2012some}
D.~Crundall, P.~Chapman, S.~Trawley, L.~Collins, E.~Van~Loon, B.~Andrews, and
  G.~Underwood, ``Some hazards are more attractive than others: Drivers of
  varying experience respond differently to different types of hazard,''
  \emph{Accident Analysis \& Prevention}, vol.~45, pp. 600--609, 2012.

\bibitem{taheri2017virtual}
S.~M. Taheri, K.~Matsushita, M.~Sasaki \emph{et~al.}, ``Virtual reality driving
  simulation for measuring driver behavior and characteristics,'' \emph{Journal
  of transportation technologies}, vol.~7, no.~02, p. 123, 2017.

\bibitem{goedicke2018vr}
D.~Goedicke, J.~Li, V.~Evers, and W.~Ju, ``Vr-oom: Virtual reality on-road
  driving simulation,'' in \emph{Proceedings of the 2018 CHI Conference on
  Human Factors in Computing Systems}, 2018, pp. 1--11.

\bibitem{xu2020much}
J.~Xu and A.~Howard, ``How much do you trust your self-driving car? exploring
  human-robot trust in high-risk scenarios,'' in \emph{2020 IEEE International
  Conference on Systems, Man, and Cybernetics (SMC)}.\hskip 1em plus 0.5em
  minus 0.4em\relax IEEE, 2020, pp. 4273--4280.

\bibitem{comparing2018Lhemedu}
Q.~Lhemedu-Steinke, G.~Meixner, and M.~Weber, ``Comparing vr display with
  conventional displays for user evaluation experiences,'' in \emph{2018 IEEE
  Conference on Virtual Reality and 3D User Interfaces (VR)}, 2018, pp.
  583--584.

\bibitem{Comparing2018Pallavicini}
\BIBentryALTinterwordspacing
F.~Pallavicini and A.~Pepe, ``Comparing player experience in video games played
  in virtual reality or on desktop displays: Immersion, flow, and positive
  emotions,'' in \emph{Extended Abstracts of the Annual Symposium on
  Computer-Human Interaction in Play Companion Extended Abstracts}, ser. CHI
  PLAY '19 Extended Abstracts.\hskip 1em plus 0.5em minus 0.4em\relax New York,
  NY, USA: Association for Computing Machinery, 2019, p. 195–210. [Online].
  Available: \url{https://doi.org/10.1145/3341215.3355736}
\BIBentrySTDinterwordspacing

\bibitem{williams2019reality}
T.~Williams, D.~Szafir, and T.~Chakraborti, ``The reality-virtuality
  interaction cube: a framework for conceptualizing mixed-reality interaction
  design elements for hri,'' in \emph{2019 14th ACM/IEEE International
  Conference on Human-Robot Interaction (HRI)}.\hskip 1em plus 0.5em minus
  0.4em\relax IEEE, 2019, pp. 520--521.

\bibitem{whitney2018ros}
D.~Whitney, E.~Rosen, D.~Ullman, E.~Phillips, and S.~Tellex, ``Ros reality: A
  virtual reality framework using consumer-grade hardware for ros-enabled
  robots,'' in \emph{2018 IEEE/RSJ International Conference on Intelligent
  Robots and Systems (IROS)}.\hskip 1em plus 0.5em minus 0.4em\relax IEEE,
  2018, pp. 1--9.

\bibitem{novitzky2020virtual}
M.~Novitzky, R.~Semmens, N.~H. Franck, C.~M. Chewar, and C.~Korpela, ``Virtual
  reality for immersive human machine teaming with vehicles,'' in
  \emph{International Conference on Human-Computer Interaction}.\hskip 1em plus
  0.5em minus 0.4em\relax Springer, 2020, pp. 575--590.

\bibitem{vailland2020vestibular}
G.~Vailland, Y.~Gaffary, L.~Devigne, V.~Gouranton, B.~Arnaldi, and M.~Babel,
  ``Vestibular feedback on a virtual reality wheelchair driving simulator: A
  pilot study,'' in \emph{Proceedings of the 2020 ACM/IEEE International
  Conference on Human-Robot Interaction}, 2020, pp. 171--179.

\bibitem{duguleana2011evaluating}
M.~Duguleana, F.~G. Barbuceanu, and G.~Mogan, ``Evaluating human-robot
  interaction during a manipulation experiment conducted in immersive virtual
  reality,'' in \emph{International Conference on Virtual and Mixed
  Reality}.\hskip 1em plus 0.5em minus 0.4em\relax Springer, 2011, pp.
  164--173.

\bibitem{naceri2019towards}
A.~Naceri, D.~Mazzanti, J.~Bimbo, D.~Prattichizzo, D.~G. Caldwell, L.~S.
  Mattos, and N.~Deshpande, ``Towards a virtual reality interface for remote
  robotic teleoperation,'' in \emph{2019 19th International Conference on
  Advanced Robotics (ICAR)}.\hskip 1em plus 0.5em minus 0.4em\relax IEEE, 2019,
  pp. 284--289.

\bibitem{taylor2020diminished}
A.~V. Taylor, A.~Matsumoto, E.~J. Carter, A.~Plopski, and H.~Admoni,
  ``Diminished reality for close quarters robotic telemanipulation,'' in
  \emph{2020 IEEE/RSJ International Conference on Intelligent Robots and
  Systems (IROS)}.\hskip 1em plus 0.5em minus 0.4em\relax IEEE, 2020, pp.
  11\,531--11\,538.

\bibitem{williams2018augmented}
T.~Williams, N.~Tran, J.~Rands, and N.~T. Dantam, ``Augmented, mixed, and
  virtual reality enabling of robot deixis,'' in \emph{International Conference
  on Virtual, Augmented and Mixed Reality}.\hskip 1em plus 0.5em minus
  0.4em\relax Springer, 2018, pp. 257--275.

\bibitem{carla}
A.~Dosovitskiy, G.~Ros, F.~Codevilla, A.~Lopez, and V.~Koltun, ``Carla: An open
  urban driving simulator,'' in \emph{Conference on robot learning}.\hskip 1em
  plus 0.5em minus 0.4em\relax PMLR, 2017, pp. 1--16.

\bibitem{slob2008state}
J.~Slob, ``State-of-the-art driving simulators, a literature survey,''
  \emph{DCT report}, vol. 107, 2008.

\bibitem{caird2011twelve}
J.~K. Caird and W.~J. Horrey, ``Twelve practical and useful questions about
  driving simulation,'' \emph{Handbook of driving simulation for engineering,
  medicine, and psychology}, vol.~2, 2011.

\bibitem{wynne2019systematic}
R.~A. Wynne, V.~Beanland, and P.~M. Salmon, ``Systematic review of driving
  simulator validation studies,'' \emph{Safety science}, vol. 117, pp.
  138--151, 2019.

\bibitem{artz2001design}
B.~Artz, L.~Cathey, P.~Grant, D.~Houston, and J.~Greenberg, ``The design and
  construction of the visual subsystem for virttex, the driving simulator at
  the ford research laboratories,'' in \emph{DSC 2001: driving simulation
  conference (Sophia Antipolis, 5-7 September 2001)}, 2001, pp. 255--262.

\bibitem{pai2019promise}
G.~Pai~Mangalore, Y.~Ebadi, S.~Samuel, M.~A. Knodler, and D.~L. Fisher, ``The
  promise of virtual reality headsets: Can they be used to measure accurately
  drivers’ hazard anticipation performance?'' \emph{Transportation research
  record}, vol. 2673, no.~10, pp. 455--464, 2019.

\bibitem{shah2018airsim}
S.~Shah, D.~Dey, C.~Lovett, and A.~Kapoor, ``Airsim: High-fidelity visual and
  physical simulation for autonomous vehicles,'' in \emph{Field and service
  robotics}.\hskip 1em plus 0.5em minus 0.4em\relax Springer, 2018, pp.
  621--635.

\bibitem{deepdrive}
``\href{https://deepdrive.io/}{DeepDrive},'' https://deepdrive.io/, [Online;
  accessed 7-Oct-2021].

\bibitem{wymann2000torcs}
B.~Wymann, E.~Espi{\'e}, C.~Guionneau, C.~Dimitrakakis, R.~Coulom, and
  A.~Sumner, ``Torcs, the open racing car simulator,'' \emph{Software available
  at http://torcs. sourceforge. net}, vol.~4, no.~6, p.~2, 2000.

\bibitem{michalik2021developing}
D.~Michal{\'\i}k, M.~Jirgl, J.~Arm, and P.~Fiedler, ``Developing an unreal
  engine 4-based vehicle driving simulator applicable in driver behavior
  analysis—a technical perspective,'' \emph{Safety}, vol.~7, no.~2, p.~25,
  2021.

\bibitem{najm2007pre}
W.~G. Najm, J.~D. Smith, M.~Yanagisawa \emph{et~al.}, ``Pre-crash scenario
  typology for crash avoidance research,'' United States. National Highway
  Traffic Safety Administration, Tech. Rep., 2007.

\end{thebibliography}


\end{document}